\documentstyle[12pt,epsf]{article}
\begin{document}
\begin{titlepage}
\title{Polarization in Top Quark Pair
Production\\
near Threshold
\footnote{Work supported in part
by KBN grant 2P30225206, by BMFT contract 056KA93P
and by DFG contract 436POL173193S.}
\footnote{The complete paper, including figures, is
also available via anonymous ftp at
ttpux2.physik.uni-karlsruhe.de (129.13.102.139) as /ttp94-28/ttp94-28.ps,
or via www at http://ttpux2.physik.uni-karlsruhe.de/preprints.html/}
}
\author{ \\
R. Harlander$^a$,
M. Je\.zabek$^{a,b}$,
J.H. K\"uhn$^a$ and T. Teubner$^a$\\
{\normalsize \it $^a$ Institut f\"ur Theoretische Teilchenphysik,
D-76128 Karlsruhe, Germany}\\
{\normalsize \it $^{b}$ Institute of Nuclear Physics,
Kawiory 26a, PL-30055 Cracow, Poland}
}
\date{}
\maketitle
\thispagestyle{empty}
\vspace{-3.6truein}
\begin{flushright}
{\bf TTP 94-28}\\
{\bf November 1994}
\end{flushright}
\vspace{3.6truein}
\begin{abstract}
\noindent
{\small
The polarization dependent momentum distributions
of top quarks and their decay products are calculated
for $t\bar t$ production at future $e^+e^-$ colliders
with polarized beams. The Green function formalism
is applied to this reaction near energy threshold.
The Lippmann--Schwinger equations for the $S$-wave and
$P$-wave Green functions are solved numerically for the
QCD chromostatic potential given by the two-loop formula
at large momentum transfers and Richardson ansatz
at intermediate and small ones.
It is demonstrated that for the longitudinally polarized
electron beam an optimally polarized sample of top quarks can
be produced.
}
\end{abstract}
\end{titlepage}
\par\noindent{\large\em Introduction}\par\noindent
There is no doubt that precise studies of top quark
interactions will lead to profound progress in particle
physics. The top quark is the heaviest fermion of the
Standard Model. Its large mass allows to probe deeply
into the QCD potential for nonrelativistic $t\bar t$ system
produced near threshold. Such a system will provide
a unique opportunity for a variety of novel QCD studies.
Polarization studies for $t\bar t$ pairs near threshold
are free from hadronization ambiguities. This is due to the
short lifetime of the top quark. The lifetime of the top
quark is shorter than the formation time of top mesons
and toponium resonances. Therefore top decays intercept
the process of hadronization at an early stage and
practically eliminate associated nonperturbative effects.

The analysis of polarized top quarks and their decays
has recently attracted considerable attention
(See \cite{Kuehn3,teupitz} and references
cited therein).
The reason is that this analysis
will result in determination of the top
quark coupling to the $W$ and $Z$ bosons either
confirming the predictions
of the Standard Model
or providing clues for physics beyond.
The latter possibility is particularly intriguing
because $m_t$
plays an exceptional role in the fermion mass spectrum.

A number of mechanisms have been suggested that will
lead to polarized top quarks. However, studies at
a linear electron-positron collider are particularly
clean for precision tests.
Moreover, close to threshold
and with longitudinally polarized electrons
one can study decays of polarized top quarks
under particularly convenient conditions:
large event rates, well-identified rest frame of the top quark,
and large degree of polarization.
At the same time,
thanks to the spectacular success
of the polarization program at SLC \cite{woods},
the longitudinal polarization of the electron beam
will be an obvious option
for a future linear collider\footnote{Another proposed
and closely related
facility is a photon linear collider.
At such a machine the high energy photon beams
can be generated via Compton scattering of laser
light on  electrons accelerated in the linac.
The threshold behaviour of the reaction
$\gamma\gamma\to t\bar t$
has been reviewed in \cite{DESYWac} and the
top quark polarization in this reaction
has been recently considered in \cite{fkk}.}.

In the present article\footnote{Some results
of these studies have been
presented in \cite{hjkt}.}
top quark polarization  is studied
in the reaction $e^+e^-\to t\bar t$.
The calculations of top polarization well above the threshold
for $t\bar t$ production were published long ago \cite{krz}.
The threshold region, however, deserves a special study.
To demonstrate that this polarization is indeed
experimentally accessible
we review briefly polarization effects in top decays.
Subsequently
we discuss the dependence of
the top quark polarization  on the longitudinal
polarizations of the beams.
Due to restricted phase space the amplitude is
dominantly $S$ wave and the electron and positron
polarizations are directly transferred to the top quark.
For a quantitative study
this simple picture
has to be extended and the modifications
originating from $S-P$ wave interference
should be taken into account.
As a consequence of final state interaction between
$t$ and $\bar t$, which leads to the familiar
QCD potential, the parton model prediction
where these terms are simply proportional to
$\beta=\sqrt{1-4m_t^2/s}$ has to be modified.
The interference terms are therefore calculated
from numerical solutions of Lippmann-Schwinger
equations.

\vskip0.5cm\noindent
{\large\em Decays of polarized top quarks}
\par\noindent
The polarization four-vector $s^\mu$
of the top quark can be determined
from the angular-energy distributions of the charged leptons
in semileptonic $t$ decays.
In the $t$ quark rest frame this distribution
is in Born approximation
the product of the energy
and the angular distributions \cite{JK89b}:
\begin{equation}
{ {\rm d}^2 \Gamma\over{\rm d}E_\ell\,{\rm d}\cos\theta} =
{1\over 2}\, \left[\, 1\,+
\,S\cos\theta \right]\,{{\rm d}\Gamma\over{\rm d}E_\ell}
\label{eq:elec1}
\end{equation}
where  $s^\mu=(0,\vec s\,)$, $S= |\,\vec s\,|$
and $\theta$ is the
angle between $\vec s$ and the direction of the charged lepton.
QCD corrections essentially do not spoil factorization~\cite{CJK91}.
Thus, the polarization analyzing power of
the charged lepton energy-angular distribution
remains maximal.
There is no factorization
for the neutrino energy-angular distribution
which is therefore less sensitive to the
polarization of the decaying top quark.
On the other hand it has been shown \cite{JK94} that
the angular-energy distribution of neutrinos
from the polarized top quark decay will allow for a particularly
sensitive test of the V-A structure of the weak charged current.

\vskip0.5cm\noindent
{\large\em Top polarization at a linear collider}
\par\noindent
We adopt the conventions of ref.\cite{alexan}
and describe
the longitudinal polarization of the $e^+e^-$ system
in its center-of-mass frame as a function of the variable
\begin{equation}
\chi = {P_{e^+}-P_{e^-}\over 1 - P_{e^+}P_{e^-}}
\end{equation}
where $P_{{e^\pm}}$ denote the polarizations of $e^\pm$
with respect to the directions of $e^+$ and $e^-$ beams,
respectively\footnote{It is conceivable
that for a future linear $e^+e^-$ collider
$P_{e^+}=0$, $P_{e^-}\ne 0$ and then $\chi = -P_{e^-}$.}.
A righthanded system of coordinates is defined through the
triplet of othogonal unit vectors: $\hat n_{\bot}$, $\hat n_{_N}$
and $\hat n_{^\|}$ where
$\hat n_{^\|}$ points in the direction of $e^-$ beam,
$\hat n_{_N}\sim \vec p_{e^-}\times \vec p_t$  is normal
to the production plane and
$\hat n_{\bot}=\hat n_{_N}\times\hat n_{^\|}$.
In the absence of phases from final state interaction,
which can be induced by higher orders in $\alpha_s$ and will be
considered elsewhere, the top quark polarization vector
is in the production plane and $P_{^\|}$ and $P_{\bot}$ denote
its longitudinal and transverse components.
The definition of $P_{^\|}$ and $P_{\bot}$ with respect
to the beam direction is convenient for the treatment
close to threshold and differs from the definition
of \cite{krz} where the quantities have been defined with respect
to the direction of flight of the top quark.
The angle $\vartheta$
denotes the angle between $e^-$ and the top quark.

\noindent
In the threshold region the top quark is nonrelativistic (with
velocity $\beta=p/m_t\sim\alpha_s$)
and the kinetic energy of the $t\bar t$ system
$E=\sqrt{s}-2m_t$ is of order ${\cal O}(\beta^2)$.
As a consequence of the $t\bar t$ interaction the top quarks
will exhibit a momentum spread (Fermi motion) even for fixed
energy $E$ \cite{jk88}. Also the polarization and the angular
distribution will depend on both $E$ and $p$.
Retaining only the terms up to
${\cal O}(\beta)$ one derives the following expressions for
the components of the polarization vector, as functions
of $E$, $p$ and $\vartheta$
\begin{eqnarray}
{\cal P}_{^\|}(p,E,\vartheta) &=& C^0_{^\|}(\chi)
+ C^1_{^\|}(\chi)\varphi(p,E)\cos\vartheta
\label{eq:2}\\
{\cal P}_\bot(p,E,\vartheta) &=&C_\bot(\chi)\varphi(p,E)\sin\vartheta
\label{eq:3}
\end{eqnarray}
and, after the integration over top quark momentum $p$,
\begin{eqnarray}
P_{^\|}(E,\vartheta)  &=& C^0_{^\|}(\chi)
+ C^1_{^\|}(\chi)\Phi(E)\cos\vartheta
\label{eq:2a}\\
P_\bot(E,\vartheta)  &=&C_\bot(\chi)\Phi(E)\sin\vartheta
\label{eq:3a}
\end{eqnarray}
where
\begin{eqnarray}
C^0_{^\|}(\chi) &=& - {a_2 + \chi a_1 \over a_1 + \chi a_2}
\\
C^1_{^\|}(\chi) &=& - (1 - \chi^2) {a_1 a_4 - a_2 a_3 \over
(a_1 +  \chi a_2)^2}  \\
C_\bot(\chi) &=& - {1\over 2} {a_4 + \chi a_3 \over a_1 + \chi a_2}
\end{eqnarray}
with
\begin{eqnarray}
a_1 &=& q_e^2 q_t^2 +  (v_e^2 + a_e^2) v_t^2 d^2
+ 2 q_e q_t   v_e v_t d \nonumber \\
a_2 &=& 2 v_e a_e v_t^2 d^2 + 2 q_e q_t  a_e v_t d \nonumber \\
a_3 &=& 4 v_e a_e v_t a_t d^2 + 2 q_e q_t a_e a_t d  \nonumber \\
a_4 &=& 2 (v_e^2 + a_e^2) v_t a_t d^2 + 2 q_e q_t  v_e a_t d
\nonumber\\
d &=& {1\over 16 \sin^2\theta_W\cos^2\theta_W}\,{s\over s - M_Z^2}
\nonumber
\end{eqnarray}
and
\vskip0.3cm\noindent
\begin{tabular}{lll}
$q_e = -1$ \qquad\qquad  & $v_e = -1 + 4 \sin^2\theta_W$ \qquad\qquad &
$a_e = -1$    \\
$q_t = {2\over 3}$  &
$v_t = 1 - {8\over 3} \sin^2\theta_W$ \qquad &
$a_t = 1 $ \\
\end{tabular}
\vskip0.3cm
\par
\begin{figure}[htb]
\epsfxsize=5.5truein
\leavevmode
\epsffile[30 300 525 515]{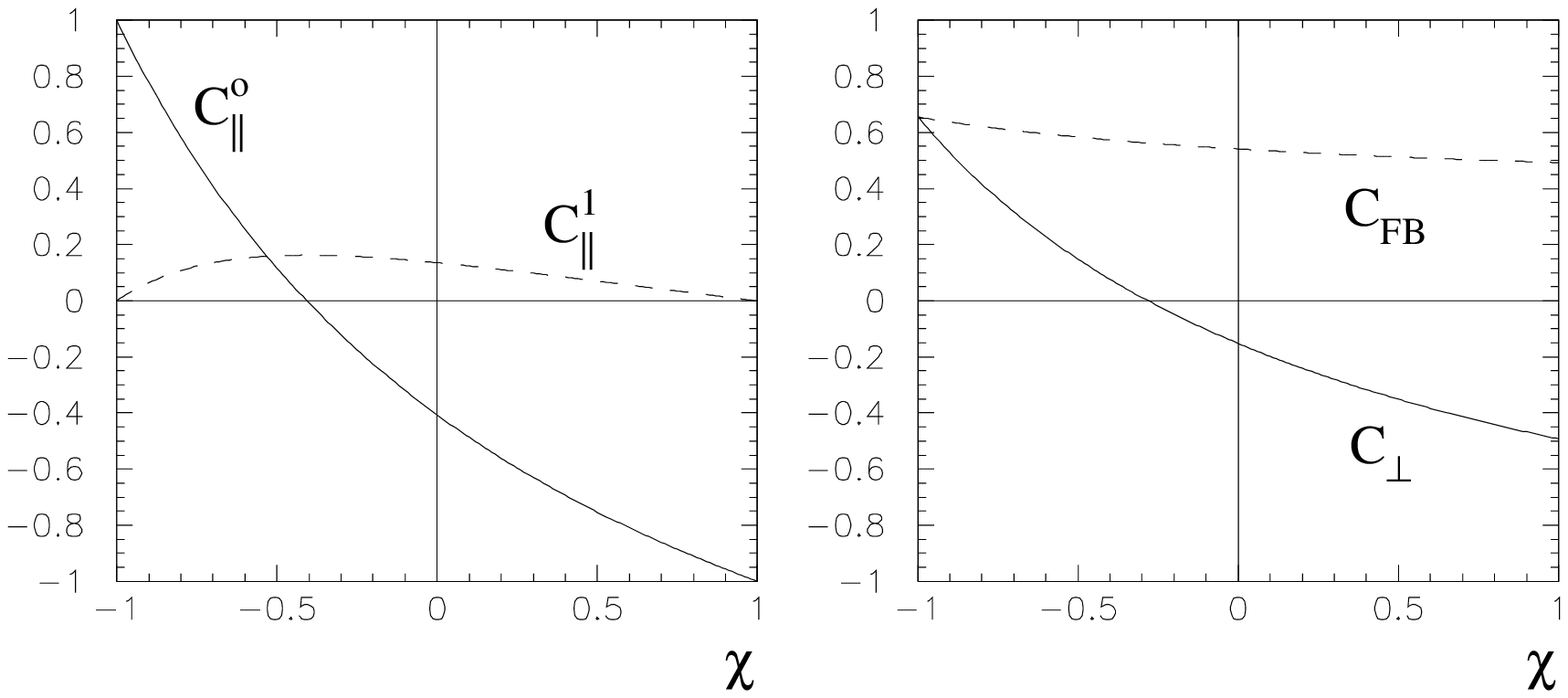}
\vskip-0.2cm
\caption{Coefficient functions: a) $C^0_{^\|}(\chi)$ -- solid line
and $C^1_{^\|}(\chi)$ -- dashed line, b) $C_\bot(\chi)$ -- solid line
and $C_{FB}(\chi)$ -- dashed line.}
\label{fig-CLchi}
\end{figure}

\noindent
The coefficients $C^0_{^\|}(\chi)$, $C^1_{^\|}(\chi)$
and $C_\bot(\chi)$ depend on the polarization $\chi$, the
electroweak coupling constants, the $Z$ mass and the center-of-mass
energy $\sqrt{s}\approx 2m_t$. They are plotted in
Fig.\ref{fig-CLchi} for $m_t=174$~GeV,
$\sin^2\theta_W = 0.2317$ and $M_Z = 91.1888\; GeV$.
$C^0_{^\|}(\chi)$ and $C^1_{^\|}(\chi)$
are shown in Fig.\ref{fig-CLchi}a as the solid and the dashed
lines, respectively, and $C_\bot(\chi)$ as the solid line in
Fig.\ref{fig-CLchi}b.
Eqs.(\ref{eq:2}) and (\ref{eq:3})
extend the results of \cite{krz,AS1} into the threshold region.

\noindent
The functions $\varphi(p,E)$  and $\Phi(E)$
in eqs.(\ref{eq:2})--(\ref{eq:3a}) replace the simple
factor $\beta$ in a calculation for free, noninteracting quarks.
They describe the complicated dynamics of the
$t\bar t$ system near threshold which in particular
includes effects of the would-be toponium resonances
and Coulomb enhancement. Nevertheless, it is possible to calculate
these functions using the Green function method.
The same functions $\varphi(p,E)$ and
$\Phi(E)$ also govern the angular distributions of the top quark
in $e^+e^-\to t\bar t$.
For fixed $E$ and $p$ the angular distribution
is given by
\begin{equation}
{{\rm d}{\cal N}\over{\rm d}\cos\vartheta }(p,E) \sim
\left( a_1 + \chi a_2 \right)\, +\,
\left( a_3 + \chi a_4 \right)\,\varphi(p,E)\cos\vartheta
\label{eq:angdis}
\end{equation}
and the corresponding forward-backward asymmetry reads
\begin{equation}
{\cal A}_{FB}(p,E) \,=\, \int^1_0 {\rm d}\cos\vartheta
{{\rm d}{\cal N}\over{\rm d}\cos\vartheta } -
 \int^0_{-1} {\rm d}\cos\vartheta
{{\rm d}{\cal N}\over{\rm d}\cos\vartheta }
\,=\,
C_{FB}(\chi)\,\varphi(p,E)
\label{eq:4}
\end{equation}
The function
\begin{equation}
C_{FB}(\chi) = {1\over 2} {a_3 + \chi a_4 \over a_1 + \chi a_2}
\end{equation}
is shown as the dashed line in Fig.\ref{fig-CLchi}b.
Integrating the distribution (\ref{eq:angdis}) over $p$
one obtaines the forward-backward asymmetry
\begin{equation}
A_{FB}(E) = C_{FB}(\chi)\Phi(E)
\label{eq:4a}
\end{equation}
which has been first obtained in \cite{Sumino2} for $\chi=0$.

\vskip0.5cm\noindent
{\large\em Lippmann-Schwinger equations}
\par\noindent
To evaluate the functions $\varphi(p,E)$ and
$\Phi(E)$ the Green function method
is adequate which
has become a standard tool for studying
$e^+e^-$ annihilation in the threshold region
\cite{FK,SP,Sumino1,JKT}.
We follow the momentum space approach \cite{JKT}
and solve the Lippmann-Schwinger
equations numerically for the $S$-wave
and $P$-wave Green functions
\begin{eqnarray}
G(p,E) &=&
G_0(p,E) +
G_0(p,E)
\int {{\rm d}^3q\over(2\pi)^3}
\tilde V\left(\,|\,\vec{p}-\vec{q}\,|\,\right)
G(q,E)
\label{eq:LS1}\\
F(p,E) &=&
G_0(p,E) +
G_0(p,E)
\int {{\rm d}^3q\over(2\pi)^3}
{{\vec p}\cdot{\vec q}\over p^2}
\tilde V\left(\,|\,\vec{p}-\vec{q}\,|\,\right)
F(q,E)
\label{eq:LS2}
\end{eqnarray}
where $p=|\,\vec p\,|$ is the momentum
of the top quark in $t\bar t$ rest frame,
\begin{equation}
G_0(p,E) = \left(\,
E- {p^2/ m_t}+ {\rm i}\Gamma_t\,\right)^{-1}
\label{eq:8}
\end{equation}
and $\Gamma_t$ denotes the top width.
The QCD potential in momentum
space $\tilde V(p)$ is described in \cite{JT}.

\begin{figure}[t]
\begin{center}
\epsfxsize=2.8in
\vskip-1cm
\leavevmode
\epsffile[130 80 440 730]{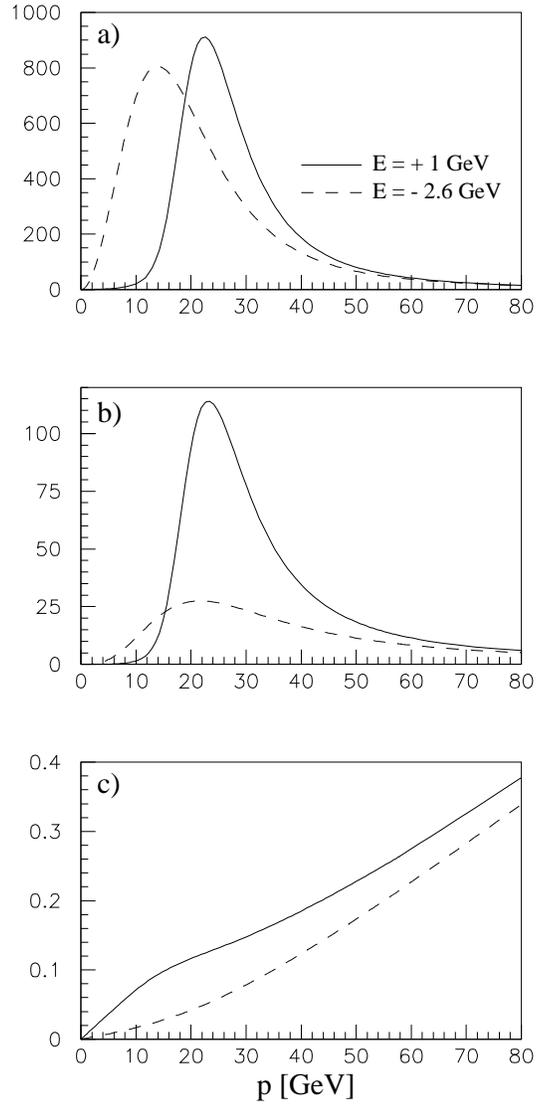}
\end{center}
\vskip-0.5cm
\caption{Momentum distributions:
a) ${\cal D}_{S-S}(p,E)$,
b) ${\cal D}_{S-P}(p,E)$, and
c) their ratio $\varphi(p,E)$
for $E= 1$ and $E= -2.6$~GeV -- solid/dashed
lines,
$m_t=174$~GeV and $\alpha_s(m_Z)=0.12$.}
\label{fig-FG}
\end{figure}

\begin{figure}[htb]
\begin{center}
\epsfxsize=3.5in
\vskip-0.1cm
\leavevmode
\epsffile[70 125 500 700]{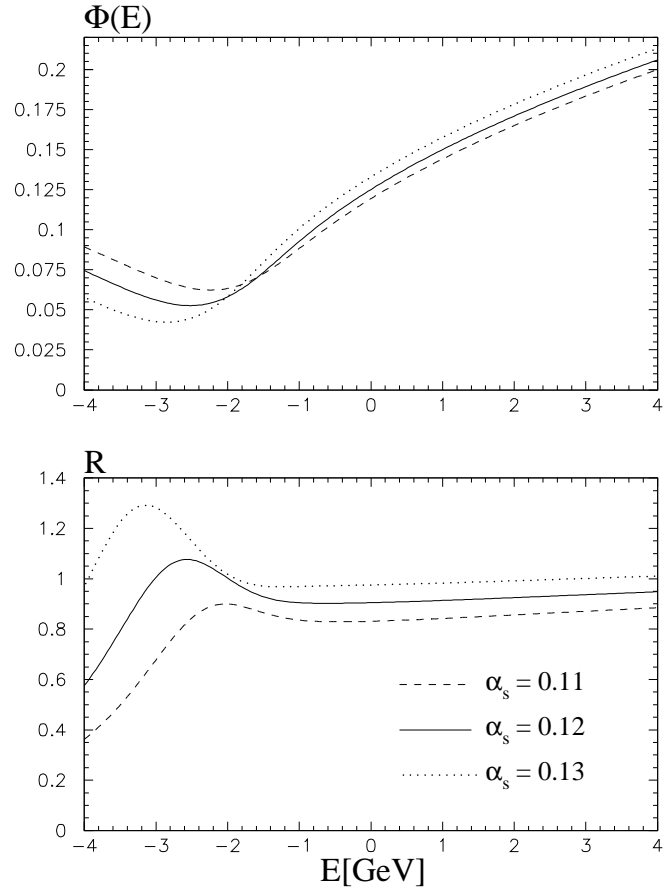}
\end{center}
\vskip-0.5cm
\caption{Energy dependence in the threshold region
of: a) $\Phi(E)$ and
b) $R=\sigma(e^+e^-\to t\bar t\,)/\sigma(e^+e^-\to \mu^+\mu^-)$
for $m_t=174$~GeV and $\alpha_s(m_Z)=0.11$, $0.12$ and $0.13$
-- dashed, solid and dotted lines, respectively.}
\label{fig-FiR}
\end{figure}

In the nonrelativistic approximation the momentum distribution
of the top quark is dominated by the $S$-wave contribution and
it is proportional to
\begin{equation}
{\cal D}_{S-S}(p,E)=p^2 \left|\, G\,\right|^2
\end{equation}
which is shown in Fig.\ref{fig-FG}a. For $E= -2.6$~GeV (dashed line)
${\cal D}_{S-S}$ is relatively wide and exhibits the shape
expected for the ground state wave function. For $E= 1$~GeV
(solid line) the distribution becomes narrower and is peaked
at a nonvanishing momentum corresponding to open top production.
Contributions of ${\cal O}(\beta)$ to
${\cal P}_{^\|}$ and ${\cal P}_{\bot}$
as well as to the forward-backward asymmetry ${\cal A}_{FB}$
arise from the interference of $S$- and $P$- waves
and are therefore proportional to
\begin{equation}
{\cal D}_{S-P}(p,E)= p^3\, {\cal R}e \left( \,G\,F^*\,\right)/m_t
\end{equation}
This distribution, shown in Fig.\ref{fig-FG}b
exhibits a significantly wider spread in momentum,
a feature naturally explained by the larger momentum
of $P$-states. The angular distribution (\ref{eq:angdis}),
the forward-backward asymmetry (\ref{eq:4}), and the components
(\ref{eq:2}) and (\ref{eq:3}) of the polarization vector
for fixed $E$ and $p$ are then governed by the ratio
\begin{equation}
\varphi(p,E)\, =\,  {
\left( 1- {4\alpha_s\over 3\pi} \right)\, {\cal D}_{S-P}(p,E)
\over
\left( 1- {8\alpha_s\over 3\pi} \right)
{\cal D}_{S-S}(p,E)  }
\end{equation}
which is plotted in  Fig.\ref{fig-FG}c.
The factors
$\left( 1- {4\alpha_s\over 3\pi} \right)$
and
$\left( 1- {8\alpha_s\over 3\pi} \right)$
originate from transversal gluon exchange  \cite{kz}.
In a similar way the function
\begin{equation}
\Phi(E) = {
\left( 1- {4\alpha_s\over 3\pi} \right)
\int_0^{p_{m}}\,{\rm d}p\, {\cal D}_{S-P}(p,E)
\over
\left( 1- {8\alpha_s\over 3\pi} \right)
\int_0^{p_{m}}\,{\rm d}p\, {\cal D}_{S-S}(p,E) }
\label{eq:Phi}
\end{equation}
describes the integrated quantities.
The upper limit
$p_{m}$ has been introduced in order to cut off
the logarithmic divergence of the numerator.
The denominator remains finite for $p_{m}\rightarrow\infty$.
In experimental analyses the
contributions of very large intrinsic momenta
will be automatically suppressed
by the separation of $t\bar t$ events from the background.
In our calculation we use $p_{m}= m_t/3$.
%
The function
$\Phi(E)$ is plotted in Fig.\ref{fig-FiR}a for $m_t=174$~GeV.
The QCD potential depends on $\alpha_s(m_Z)$ and our results
are presented for $\alpha_s=0.11\,/\,0.12\,/\,0.13$
as dotted/solid/dashed lines.
For a comparison in
Fig.\ref{fig-FiR}b the annihilation cross section
$\sigma(e^+e^-\to t\bar t\,)$
is shown in units of
$\sigma(e^+e^-\to \mu^+\mu^-)$.
The increase of $\Phi(E)$ with $E$ resembles qualitatively the
behaviour $\beta\approx \sqrt{E/m_t}$ expected for noninteracting
quarks. Linear terms in $\beta$ amount to typically 0.1 and could,
therefore, be accessible experimentally, whereas relativistic
corrections $\sim\beta^2$ are evidently  small.

\par\noindent
Combining Figs.(\ref{fig-CLchi})--(\ref{fig-FG})
the predictions for the polarizations
and the angular distributions can be discussed easily.
The coefficient $C^0_{^\|}(\chi)$
governs the leading term for the longitudinal polarization.
(For electroweak corrections to this quantity see
\cite{GKKS,GK}.)
For $\chi=\pm 1$ the top quark polarization is simply given by
the polarization of the incoming electrons, resulting in
$C^0_{^\|}(\chi)=\mp 1$ and
$C^1_{^\|}(\chi)=0$, and consequently in a nearly completely
polarized sample of top quarks, {\em a strong argument for top
production with polarized beams}. The deviation from the $\beta=0$
prediction for the longitudinal top polarization is maximal
at $\chi\approx -0.4$, where
$C^0_{^\|}(\chi)$ happens to vanish, but remains generally small.
The coefficient
$C_\bot(\chi)$ is small for unpolarized beams. For
$\chi= -1$ however, it amounts to 0.65 which leads to sizeable
polarization of top quarks perpendicular to the beam,
if they are produced at large angles.
The angular distribution and the forward-backward asymmetry $A_{FB}$,
on the other hand, are fairly insensitive towards beam polarization
in the threshold region. They can be used for experimental
determination of $\Phi(E)$ which will test the QCD
description of $t\bar t$
interactions and may lead to an independent
determination of $m_t$ and $\alpha_s$.

\vskip0.5cm\noindent
{\large\em Summary}
\par\noindent
Top quark polarization can be reliably predicted in the threshold
region. Employing longitudinally polarized beams an optimally
polarized sample of top quarks may be produced, allowing for
precision studies of $tbW$ coupling. Small corrections of order
$\beta$ arise from the admixture of $P$ waves. These can be
calculated using Green function techniques. The predictions
are sensitive towards the potential model, leading to additional
constraints on the value of $\alpha_s$.


\end{document}